\documentclass[utf8]{frontiersFPHY} 

\usepackage{url,hyperref,lineno,microtype,subcaption}
\usepackage[onehalfspacing]{setspace}

\def\keyFont{\fontsize{8}{11}\helveticabold }
\def\firstAuthorLast{Zhang {et~al.}} 
\def\Authors{Jian-Fu Zhang\,$^{1,2,*}$, Ru-Yue Wang\,$^{1}$}
\def\Address{$^{1}$Department of Physics, Xiangtan University, Xiangtan, Hunan 411105, China \\
$^{2}$Key Laboratory of Stars and Interstellar Medium, Xiangtan University, Xiangtan, Hunan 411105, China  }
\def\corrAuthor{Jianfu Zhang}

\def\corrEmail{jfzhang@xtu.edu.cn}

\begin{document}
\firstpage{1}

\title[Measurement of MHD turbulence properties]{Measurement of MHD turbulence properties by synchrotron radiation techniques} 

\author[\firstAuthorLast ]{\Authors} 
\address{} 
\correspondance{} 

\extraAuth{}

\maketitle

\begin{abstract}

It is well-known that magnetohydrodynamic (MHD) turbulence is ubiquitous in astrophysical environments. The correct understanding of the fundamental properties of MHD turbulence is a prerequisite for revealing many key astrophysical processes. The development of observation-based measurement techniques has significantly promoted MHD turbulence theory and its implications in astrophysics. After describing the modern understanding of MHD turbulence 
based on theoretical analysis and direct numerical simulations, we review recent developments related to synchrotron fluctuation techniques. Specifically, we comment on the validation of synchrotron fluctuation techniques and the measurement performance of several properties of magnetic turbulence based on data cubes from MHD turbulence simulations and observations. Furthermore, we propose to strengthen the studies of the magnetization and 3D magnetic field structure's measurements of interstellar turbulence. At the same time, we also discuss the prospects of new techniques for measuring magnetic field properties and understanding astrophysical processes, using a large number of data cubes from the Low-Frequency Array (LOFAR) and the Square Kilometre Array (SKA).

\tiny
 \keyFont{ \section{Keywords:} Magnetohydrodynamic turbulence, Polarization radiation, Interstellar magnetic fields, Statistic methods } 
\end{abstract}

\section{Introduction}
Turbulence is ubiquitous in our universe. As is perceived in everyday life, the movement of the fluid leads to the development of hydrodynamic turbulence. Unlike our terrestrial world, conductive fluids in astrophysics naturally induce magnetic field fluctuations leading to MHD turbulence due to the presence of almost all material in an ionization plasma state. 
The self-similarity over a large dynamical range of MHD turbulence in astronomical environments has been well confirmed by observations, as shown by the power-law spectrum of interstellar electron density fluctuations on the plane of the sky (Armstrong et al., \citeyear{Armstrong_etal:1995}; Chepurnov \& Lazarian, \citeyear{ChepurnovLazarian:2010}). In addition, the presence of MHD turbulence has been also pointed out by observations that far as in galaxy clusters (Zhuravleva et al., \citeyear{Zhuravleva2019}) as well as by in-situ direct measurements in the local solar wind (Bruno \& Carbone, \citeyear{Bruno2013}).

It is known that MHD turbulence plays an important role in many critical astrophysical processes, such as star formation (Mac Low \& Klessen, \citeyear{MacLow_Klessen:2004}; McKee \& Ostriker \citeyear{McKee_Ostriker2007}; Ostriker, \citeyear{Ostriker2009}; Dib et al., \citeyear{Dib2010}; Crutcher, \citeyear{Crutcher2012}), magnetic reconnection (Lazarian \& Vishniac, \citeyear{Lazarian_Vishniac1999}, henceforth LV99; Lazarian et al., \citeyear{Lazarian_etal_2020}), magnetic field amplification (Federrath, \citeyear{Federrath2016}), cosmic rays acceleration (Gaches et al., \citeyear{Gaches_etal_2021}; Zhang \& Xiang, \citeyear{Zhang_Xiang2021}) and diffusion (Yan \& Lazarian, \citeyear{Yan_Lazarian2008}; Xu \& Lazarian, \citeyear{Xu_Lazarian2018}; Lazarian \& Xu, \citeyear{Lazarian_Xu_2021}; Fornieri et al., \citeyear{Fornieri2021} for a comprehensive numerical study), and heat conduction in galaxy clusters (Yuan et al., \citeyear{Yuan2015}; Bu et al., \citeyear{BuWuYuan2016}). Naturally, a thorough understanding of MHD turbulence is essential in order to better describe a large number of astrophysical processes that operate over a large number of physical scales.

MHD turbulence with the nonlinear complexity generally exhibits diverse properties at a microscopic plasma level, but it shows regularity behind chaotic phenomena for turbulent fluctuations and allows a statistical description at a macroscopic level (Biskamp, \citeyear{Biskamp2003}). Taking statistical description into account, commonly used methods are the power spectrum, the correlation and structure functions (see Monin \& Yaglom (\citeyear{Monin_Yaglom1975}) for an old book; Boldyrev et al., \citeyear{Boldyrev2002a}; \citeyear{Boldyrev2002b}), by which one mostly considers volume and time averaging of physical variables to implement the statistical ensemble measurements of turbulence. 
In addition, the Delta-variance spectrum (Stutzki et al., \citeyear{Stutzki1998}; Dib et al., \citeyear{Dib2021}), as another manifestation of the power spectrum, has been used to characterize turbulence properties.

In recent years, numerical simulations or experiments combined with analytical theory has significantly promoted the study of MHD turbulence (Goldreich \& Sridhar, \citeyear{Goldreich_Sridhar1995}, thereafter GS95; LV99; Cho \& Lazarian, \citeyear{Cho_Lazarian2002}; Kowal \& Lazarian, \citeyear{Kowal_Lazarian2010}; Beresnyak \citeyear{Beresnyak2017}). In particular, the numerical experiments are gradually alleviating the gap between the realistic MHD turbulence physics and the analytical theory. However, due to the large-scale characteristics inherent in astrophysical systems resulting in a high Reynolds number $R_e > 10^{10}$, it is challenging to simulate a realistic astrophysical environment. The currently available 3D MHD simulations are limited by the Reynolds number of about $R_{\rm e}\simeq 10^5$ (e.g., Beresnyak, \citeyear{Beresnyak2019}). Therefore, a new observation-based research perspective is to break the bottleneck of the direct numerical simulation. 

In addition to the study of electron density information, there are two additional sources of information regarding MHD turbulence in astrophysics: turbulent velocity and magnetic fields. The former can be retrieved from spectrometric observations of emission lines (e.g., 
Lazarian et al., \citeyear{Lazarian_etal2002}; Lazarian \& Pogosyan, \citeyear{Lazarian_Pogosyan2004}), while the latter, by radio observations (Gaensler et al., \citeyear{Gaensler2011}; Lazarian \& Pogosyan, \citeyear{Lazarian_Pogosyan2012}; \citeyear{Lazarian_Pogosyan2016}, hereafter, LP12 and LP16). The main purpose of this paper is to review the progress made in measuring the fundamental properties of MHD turbulence using the radio synchrotron radiation method.

Section 2 describes the fundamentals of MHD turbulence theory necessary to understand what MHD turbulence is and develop magnetic field measurement techniques. In Section 3, we provide some basic formulae of synchrotron radiative processes in MHD turbulence. Section 4 presents various statistical techniques that are being currently developed, the feasibility of which is tested using synthetic, realistic simulation and observational data in Section 5. Finally, we give a discussion in Section 6.

\section{Fundamentals of MHD turbulence theory}\label{thoer}
\subsection{Turbulence driving mechanism} 
The turbulence energy source driving varies with different astrophysical environments. Here, to simplify considerations, we divide the driving sources into two categories: external and internal driving mechanisms. The former has been considered as a common way for turbulence driving in the simulations of interstellar turbulence, due to the convenience of setting external forces. The most typical cases are supernova explosions in ISM (Spitzer, \citeyear{Spitzer_1978}; Ostriker, \citeyear{Ostriker2009}; Chamandy \& Shukurov, \citeyear{Chamandy_Shukurov2020}; Dib et al., \citeyear{Dib2006}), merger events and active galactic nuclei (AGN) outflows in the intercluster medium (Subramanian et al., \citeyear{Subramanian_etal_2006}; Chandran, \citeyear{Chandran2005}; Lazarian et al., \citeyear{Lazarian_etal_2020}) and outflow in molecular clouds (Carroll et al., \citeyear{Carroll_etal_2009}) and young stellar objects (Federrath, \citeyear{Federrath2016}). In contrast, the latter is also accepted generally, such as magnetic reconnection driving/mediating turbulence in various regimes (Kowal et al., \citeyear{Kowal_etal_2017}; Huang \& Bhattacharjee, \citeyear{Huang2016}; Cerri \& Califano, \citeyear{cerri2017}; Franci et al.,  \citeyear{Franci2017}; Dong et al., \citeyear{Dong2018}), the magnetorotational instability-generated turbulence (Gardiner, \citeyear{Gardiner_Stone_2005}; Fromang, \citeyear{Fromang_2013}; Kunz et al., \citeyear{Kunz2016}; Riols et al., \citeyear{Riols2017}; Zhdankin et al., \citeyear{Zhdankin2017}; Sellwood \& Balbus, \citeyear{Sellwood_Balbus1999}). Indeed, for realistic astrophysical processes such as molecular clouds and supernovae, their drivings may be involved in internal and external mechanisms simultaneously. 

Sometimes the complexity of turbulent driving is manifested by the coexistence of multiple driving mechanisms in the same environment. One of the most classic cases is the driving process in the ISM, where various drivers may occur simultaneously: expanding shells and gravitational instability at about 1000 pc scales arising from galaxy mergers; expanding shells, MRI and cloud collisions at about 100 pc scales from supernova explosion; and stellar feedback at 10 pc to sub-pc scales by outflow. In the turbulence research community, for simplicity, one usually uses external driving with the solenoidal or compressible approach, which is more effective at limited numerical resolution. In recent years, internal driving mechanisms have gradually become the focus of research, but only at the cost of expensive numerical resources (Beresnyak, \citeyear{Beresnyak2017}; Kowal et al., \citeyear{Kowal_etal_2017}).

\subsection{Basic properties of MHD turbulence}
In the spirit of understanding the nature of turbulence from first principles, we here provide the visco-resistive equations that govern the evolution of a magnetized plasma as follows\footnote{ These equations can be applied to describe the plasma dynamics in the case of the long wavelengths and low frequencies with respect to plasma characteristic scales and frequencies, and the associated plasma approximations such as quasi neutrality, no displacement current in Ampére's law because of non-relativistic limit, as well as resistive Ohm's law (refer to Chapter 3 of Krall \& Tivelpiece (\citeyear{krall1973}) for further details).}:
\begin{equation}
\frac{\partial \rho}{\partial t}+\nabla\cdot(\rho {\bf v})=0, \label{continuity}
\end{equation}
\begin{equation}
\rho(\frac{\partial}{\partial t}+{\bf{v}}\cdot\nabla){\bf{v}}={\frac{{\bf{j}}\times{\bf{B}}}{c}}-\nabla{{p_{\rm g}}}+\rho\nu{\nabla}^{2}{\bf{v}},\label{momentum}
\end{equation}
\begin{equation}
\frac{\partial {p_{\rm g}}}{\partial t}+{\bf v}\cdot\nabla {p_{\rm g}}+\Gamma {p_{\rm g}}\nabla\cdot {\bf v}=0, \label{pressure}
\end{equation}
\begin{equation}
\frac{\partial{\bf B}}{\partial t}-\nabla\times({\bf{v}}\times{\bf B})=\eta\nabla^{2}{\bf{B}},\label{induction}
\end{equation}
\begin{equation}
\nabla\cdot{\bf B}=0,\label{Coulomb}
\end{equation}
describing the continuity, Eq. (\ref{continuity}), the momentum conservation, Eq.(\ref{momentum}), the adiabatic invariance, Eq. (\ref{pressure}),
the magnetic induction, Eq. (\ref{induction}), and the divergence of magnetic field, Eq. (\ref{Coulomb}), respectively. Here, ${p_{\rm g}}=c_{\rm s}^{2}\rho$ is the gas pressure, $c_{\rm s}=\sqrt{\Gamma p_{\rm g}/\rho}$ the speed of sound, $t$ the evolution time of the fluid, $\nu$ the kinematic viscosity, ${\eta}=c^2/(4\pi\sigma)$ the magnetic diffusivity, $\sigma$ the conductivity, and $\Gamma$ the adiabatic index. Other physical quantities have their usual meanings. The interaction of plasma fluids and magnetic field was formulized in the above equations, by which a series of physical quantities characterizing the properties of the MHD turbulence can be better understood. 

Let us replace the operator $\nabla$ by ${\nabla}/{L}$ and set ${\bf v}=v {\bf \tilde{v}}=L/t{\bf \tilde{v}}$ in Eq. (\ref{momentum}) and ${\bf B}={B \bf \tilde{B}}$
in Eq. (\ref{induction}), where $L$ is the characteristic scale length at the typical velocity $v$, as well as $\bf \tilde{v}$ and $\bf \tilde{B}$ are dimensionless. After a simple analytical derivation (including only the viscosity term in Eq. \ref{momentum}), we obtain 
\begin{equation}
\frac{\partial{\bf \tilde{v}}}{\partial t}+{\bf \tilde{v}}\cdot{\nabla}{\bf \tilde{v}}=\frac{\nu}{vL}{\nabla}^{2}{\bf \tilde{v}},\label{momentum1}
\end{equation}

\begin{equation}
\frac{\partial{\bf \tilde{B}}}{\partial t}-{\nabla}\times({\bf \tilde{v}}\times{\bf \tilde{B}})=\frac{\eta}{Lv}\nabla^{2}{\bf \tilde{B}}.\\\
\label{induction1}
\end{equation}
Through the coefficients of the equations, we define two dimensionless physical quantities
\begin{equation}
R_{\rm e}=\frac{Lv}{\nu},\ \ \
R_{\rm m}=\frac{Lv}{\eta},
\end{equation}
where $R_{\rm e}$ is the kinetic Reynolds number characterizing the ratio of the viscous timescale to the advection timescale, and $R_{\rm m}$ is the magnetic Reynolds number denoting the ratio of resistive diffusion timescale to the advection timescale. Considering the temperature $T\simeq 10^4$ K related to $\eta\simeq 0.42\times 10^7T_{\rm eV}^{-3/2}\rm cm^2/s$, the scale length $L\simeq 1$ pc, and the velocity $v\simeq 1$ km/s, one can obtain a large $R_{\rm m}\simeq 10^{16}$ as an example of turbulence setting (e.g., Brandenburg \& Subramanian, \citeyear{Brandenburg2005}). The relative dominance between the kinematic viscosity and magnetic diffusivity is described by the magnetic Prandtl number:
\begin{equation}
P_{\rm r}=\frac{R_{\rm m}}{R_{\rm  e}}=\frac{\nu}{\eta}.
\end{equation}

Similarly, combining the adiabatic invariance, Eq. (\ref{pressure}) and momentum conservation, Eq. (\ref{momentum}), one can obtain the sonic Mach number (Biskamp, \citeyear{Biskamp2003} for a detailed derivation)
\begin{equation}
M_{\rm s}=v/c_{\rm s},
\end{equation}
for the high-$\beta$ plasma case.
This parameter can be used to describe the information of compressibility of MHD turbulence. For the low-$\beta$ plasma case, the momentum conservation Eq. (\ref{momentum}) is dominated by the magnetic pressure. 
On multiplying Eq. (\ref{momentum}) by ${\bf v}=v \bf \tilde{v}$, the quantity characterizing the magnetization strength is called the Alfv{\'e}nic Mach number
\begin{equation}
M_{\rm A}=v/V_{\rm A},
\end{equation}
where $V_{\rm A}=B/\sqrt{4\pi\rho}$ is the Alfv\'enic velocity.

Considering current density ($\nabla\times{\bf B}=\frac{4\pi}{c} {\bf j}$) related to the magnetic field by Ampere's law, the Lorentz force can be written in the following form:
\begin{equation}
\frac{{\bf j}\times{\bf B}}{c}=\frac{({\bf B}\cdot\nabla){\bf B}}{4\pi}-\nabla\frac{{\bf B^2}}{8\pi}, \label{Lorentz}
\end{equation}
where the first term on the right-hand side describes the magnetic tension force and the second term magnetic pressure force.
The plasma parameter $\beta$, denoting the ratio of gas pressure to magnetic one, is defined as
\begin{equation}
\beta=\frac{p_{\rm g}}{{B^2}/8\pi},
\end{equation} 
which can be rewritten as
\begin{equation}
\beta=\frac{2M_{\rm A}^2}{M_{\rm s}^2},
\end{equation}
with the sonic and Alfv{\'e}n Mach numbers.

As for the description of the properties of helicity in MHD turbulence, one can introduce another two quantities termed the magnetic helicity and cross helicity, which are respectively written as 
\begin{equation}
H^{M}=\int_{V}{\bf A}\cdot {\bf B} dV,
\end{equation}
\begin{equation}
H^{C}=\int_{V}{\bf v}\cdot {\bf B} dV.
\end{equation}
Here, ${\bf A}$ is the vector potential with ${\bf B}=\nabla\times {\bf A}$. The former can measure the twist and linkage of the magnetic field lines, and the latter can reflect the overall correlation of magnetic and velocity fields (Falgarone \& Passot, \citeyear{Falgarone_Passot_2003}).

\subsection{Cascade processes of MHD turbulence}

\subsubsection{Power law scaling of turbulence cascade}\label{scaling_power}

The pioneering work regarding incompressible hydrodynamic turbulence (Kolmogorov, \citeyear{Kolmogorov1941}, henceforth K41) demonstrated that the energy is injected at large scale (called the injection scale) and dissipated at small scales (the dissipation scale). The range between these two scales is called the inertial range, where the energy cascades from injection scale to dissipation scale with negligible energy losses, that is, the energy transfer rate in this range keeps a constant, 
\begin{equation}
\epsilon \thickapprox u_{l}^{2}/t_{\rm cas}=const., \label{energy_transfer}
\end{equation}
where $u_{l}$ is a characteristic velocity, and $t_{\rm cas}=l/{u_{l}}$ is the cascading timescale. This expression indirectly reflects the relationship between velocity $u_{l}$ of fluid elements and their scales $l$, namely $u_{l}\propto ({\epsilon l_\perp})^{1/3}$. Therefore, we have
the well-known Kolmogorov 1D spectrum 
\begin{equation}
E(k)\sim {\epsilon}^{2/3} k^{-5/3},
\end{equation}
where $k$ is the wave number. Alternatively, the 3D spectrum is expressed as 
\begin{equation}
E_{\rm 3D}(k)\sim k^{-11/3}.
\end{equation}
It is noted that this 3D spectrum can be transformed into the 1D one by the transformation of the volume element $d^3{\bf{k}} \sim k^2dk$, i.e., $E(k)\sim k^2 E_{\rm 3D}(k) \sim k^{-5/3}$.

For incompressible magnetic turbulence, two classical works (Iroshnikov, \citeyear{Iroshnikov1963}; Kraichnan, \citeyear{kraichnan1965inertial}, henceforth IK) introduced nonlinear energy models to study the power-law scaling from turbulence cascade. Different from the K41 energy cascade by the interaction of eddies, they considered that the energy was transferred via the Alfv{\'e}n waves along the local magnetic field. The relation between the velocity and scale is expressed as $u_{l}\propto (\epsilon V_{\rm A}l)^{1/4}$ after
considering the influence of the magnetic field. The power spectrum in the IK theory follows
\begin{equation}
E(k)\sim(\epsilon V_{\rm A})^{1/2}k^{-3/2}
\end{equation}
in the inertial range. Although this model considered the role of the magnetic field, it ignored a critical issue that turbulence should be anisotropic in the presence of the magnetic field. This limitation was pointed out by later studies (Montgomery, \citeyear{montgomery1982}; Shebalin et al., \citeyear{shebalin1983}).


With the anisotropy due to turbulent magnetic fields, GS95 studied the nonlinear energy cascade of incompressible and strong MHD turbulence. They predicted that the motions of eddies perpendicular to the magnetic field have similar properties as Kolmogorov turbulence, i.e., {$v_{\perp}\propto (\epsilon l_{\perp})^{1/3}$}. The resulting Kolmogorov-like energy spectra is given by
\begin{equation}
E(k_{\perp})\varpropto \epsilon^{2/3} k_{\perp}^{-5/3}, 
\end{equation}
where $k_{\perp}$ represents the wave-vector component perpendicular to the magnetic field. Notice that the magnetic field in the original paper refers to the global frame of reference which has been corrected as the local frame of reference by the later intensive studies (LV99; Cho \& Vishniac, \citeyear{Cho_Vishniac2000};  Maron \& Goldreich, \citeyear{Maron_Goldreich2001}; Cho \& Lazarian, \citeyear{Cho_Lazarian2002}, henceforth CL02; \citeyear{Cho_Lazarian2003}, henceforth CL03). In particular, LV99 theory pointed out that the reconnection of turbulent magnetic fields takes place within the eddy turnover time and the motions of eddies perpendicular to the magnetic field are not controlled by the magnetic field tension. The magnetic field here refers to the local magnetic field around the eddies. 

Besides, for the weak MHD turbulence, the energy transfer rate is expressed $\epsilon \propto (V_{\rm L}^{4} l_{\parallel}/V_{\rm A} l_{\perp}^{2})$, resulting in $v_{l} \propto V_{\rm L} (l_{\perp}/L_{\rm in})^{1/2}$. Thus, the power spectrum of weak turbulence follows (LV99; Galtier et al., \citeyear{Galtier2000}) 
\begin{equation}
E(k_{\perp})\sim k_{\perp}^{-2}.
\end{equation}

\subsubsection{Anisotropy}\label{anisoth}

Except for the prediction of the exponent of $-5/3$ in the perpendicular direction mentioned above, another important contribution of GS95 theory is the prediction of the scale-dependent anisotropy. They assumed that the relative motion perpendicular and parallel to the magnetic field is described by the critical balance condition, namely
$l_{\perp}^{-1}v_{\perp}\sim l_{\parallel}^{-1}V_{\rm A}$, where $v_{\perp}$ is fluctuation velocity of turbulence at the scale $l_{\perp}$, $l_{\parallel}$ and $l_{\perp}$ represent the scales parallel and perpendicular to the magnetic field, respectively. Based on the critical balance condition and the relation of {$v_{\perp}\propto(\epsilon l_{\perp})^{1/3}$}, two scales are related by
\begin{equation}
l_{\parallel}\sim V_{\rm A} \epsilon^{{-1}/{3}} l_{\perp}^{{2}/{3}},
\end{equation}
which is the so-called scale-dependent anisotropy theory. This implies that the smaller the scale, the larger the anisotropy. Note that this anisotropic relationship holds in the local frame of reference rather than in the global frame of reference. As for the latter, the  
anisotropy is scale-independent and is determined by the anisotropy of the largest eddies (CL03).

It should be stressed that GS95 focused on trans-Alfv{\'e}nic turbulence, i.e., $M_{\rm A}\approx 1$, with the turbulence kinetic energy equal to the magnetic energy. Later, the theory was generalized to sub-Alfv{\'e}nic $M_{\rm A}<1$ and super-Alfv{\'e}nic $M_{\rm A}>1$ cases, respectively (LV99; Lazarian, \citeyear{Lazarian2006}). When the turbulence is driven with the injection velocity $V_{\rm L}$ less than $V_{\rm A}$ at the injection scale $L_{\rm in}$, it initially undergoes a weak turbulence cascade process with $v_l\propto l_{\perp}^{1/2}$, where the parallel scale $l_\parallel$ does not change (LV99; Galtier et al., \citeyear{Galtier2000}). With the intensification of the interaction of wave packets, the turbulence changes to strong turbulence at the transition scale $L_{\rm tr}=L_{\rm in}M_{\rm A}^{2}$ and continues up to the dissipative scale $L_{\rm dis}$.
In the inertial range, the scales perpendicular and parallel to the magnetic field have the following relationship
\begin{equation}
l_{\parallel}\approx L_{\rm in}^{1/3}l_{\perp}^{2/3}M_{\rm A}^{-4/3}. \label{eq:lpepa}
\end{equation}
Compared with the trans-Alfv{\'e}nic turbulence, this relationship has an additional factor related to the magnetization $M_{\rm A}^{-4/3}$. In addition, the turbulent velocity is written as  
\begin{equation}
v_{l}\approx V_{\rm L}(\frac{l_{\perp}}{L_{\rm in}})^{1/3}M_{\rm A}^{1/3}, \label{eq:01}
\end{equation}
which still follows the Kolmogorov-type cascade, i.e., $v_{l}\propto l_{\perp}^{1/3}$.

In the case of the super-Alfv{\'e}nic turbulence, there is almost no constraint of the magnetic field at the injection scale $L_{\rm in}$, so the turbulence follows the law of hydrodynamic turbulence. With the development of turbulence, it experiences a transition from hydrodynamic turbulence to MHD one at the transition scale $L_{\rm A}=L_{\rm in}M_{\rm A}^{-3}$.
In the inertial range from $L_{\rm A}$ to $L_{\rm dis}$, the turbulence exhibits again an anisotropy consistent with GS95 theory, with the scale relation of 
\begin{equation}
l_{\parallel}\approx L_{\rm in}^{1/3}l_{\perp}^{2/3}M_{\rm A}^{-1}, \label{eq:02}
\end{equation}
and the velocity relation of 
\begin{equation}
v_{l}\approx V_{\rm L}(\frac{l_{\perp}}{L_{\rm in}})^{1/3}. \label{eq:01}
\end{equation}
In the strongly turbulent region, the scale-dependent anisotropy becomes more significant as the scale decreases.  

\subsection{Compressibility of MHD turbulence} \label{Compre}
Compressibility is an important addition to the MHD turbulence theory. In fact, in a real astrophysical environment such as a molecular cloud, the MHD turbulence is essentially compressible and is composed of three plasma modes, i.e., Alfv{\'e}n, slow and fast modes. The idea of the decomposition of MHD turbulence into individual modes was originally formulated in Dobrowolny et al. (\citeyear{Dobrowolny_etal_1980}), where they mainly dealt with small amplitude perturbations in incompressible MHD turbulence. 

The actual decomposition in numerical simulation has been implemented using Fourier transformation (CL02; CL03). Covering the gas-pressure-dominated and magnetic-pressure-dominated turbulence regimes, they revealed the spectrum and anisotropy in the compressible MHD turbulence. Typically, the Alfv{\'e}n mode in the high- and low-$\beta$ regimes follows the scaling slope and scale-dependent anisotropy of GS95
\begin{equation}
E^{A}(k)\propto k^{-5/3},\ \ \ k_{\parallel}\propto k_{\perp}^{2/3}. \label{alfven}
\end{equation}
The slow mode, as a passive mode, is dominated by the Alfv\'en mode, and has  properties similar to those of the Alfv\'en mode:
\begin{equation}
E^{s}(k)\propto k^{-5/3},\ \ \  k_{\parallel}\propto k_{\perp}^{2/3}. \label{slow}
\end{equation}
The fast mode, different from the Alfv{\'e}n and slow modes, possesses isotropic properties similar to acoustic turbulence, with the following scaling relations 
\begin{equation}
E^{f}(k)\propto k^{-3/2},\ \ \  k_{\parallel}\propto k_{\perp}. \label{fast}
\end{equation}

Later, Kowal \& Lazarian (\citeyear{Kowal_Lazarian2010}) has advanced the studies on mode decomposition using a wavelet method to treat the large amplitude perturbations. This method enables us to trace the direction of the local magnetic field and has significant advantages over the Fourier transform done in the frame of reference of the mean magnetic field. 

\subsection{Controversy}\label{Controv}
To explain the numerical simulations in Maron \& Goldreich (\citeyear{Maron_Goldreich2001}), several studies that have attracted community attention proposed a particular process termed as dynamical alignment to modify the GS95 spectrum from $k^{-5/3}$ to $k^{-3/2}$ (Boldyrev, \citeyear{Boldyrev2005}; Boldyrev, \citeyear{Boldyrev2006}; Mason et al., \citeyear{Mason_etal_2006}). At present, there are numerical results that are compatible with a dynamic alignment model (e.g., Chandran et al., \citeyear{Chandran2015}). Moreover, solar-wind observations seem to find some scale-dependent alignment (e.g., Wicks et al.,  \citeyear{wicks2013}) or the emergence of three-dimensional anisotropy (e.g., Chen et al., \citeyear{Chen2011}) which would also point toward a scale-dependent alignment of some sort (see also Wang et al., \citeyear{WangHe2020} for similar findings from the end of the MHD cascade to the kinetic scales).

At scales close to the dissipation scale, another modification of GS95 theory was proposed by Mallet \& Schekochihin  (\citeyear{Mallet_Schekochihin2017}) for their intermittency model and Mallet et al. (\citeyear{Mallet2017}) for their reconnection-mediated turbulence model (see also Loureiro \& Boldyrev, \citeyear{Loureiro2017}), the testing/confirming of which is difficult due to current limitations in computational abilities.

However, the anisotropy predicted in Bolryrev’s work is inconsistent with some recent numerical simulations (Beresnyak \& Lazarian, \citeyear{Beresnyak_Lazarian_2019}; Beresnyak, \citeyear{Beresnyak2019}). Many studies claimed that the deviation from the spectral index -5/3 is transient, namely localized in the vicinity of the injection scale, and does not extend to the whole inertial range (Beresnyak \& Lazarian, \citeyear{Beresnyak_Lazarian_2010}; Beresnyak, \citeyear{Beresnyak2013}; Beresnyak, \citeyear{Beresnyak2014}). To the best of my knowledge, this issue is still open. Anyway, the current attempts still do not change the paradigm of the modern MHD turbulence theory.

Synchrotron fluctuation techniques reviewed in this paper are developed closely based on the modern understanding of MHD turbulence theory (e.g., GS95; LV99; CL02) without more concern for its controversial appearance. Fortunately, modifications attempted to fundamental theory are not expected to seriously affect the measurement of the techniques. This is because the development of measurement techniques is independent of a specific scaling slope of turbulence and is constrained in the strongly turbulent range, while modifications to turbulence theory focus on the driving or dissipation scale.

\begin{figure*}[t]
\centerline{\includegraphics[width=0.98\textwidth,height=0.38\textheight]{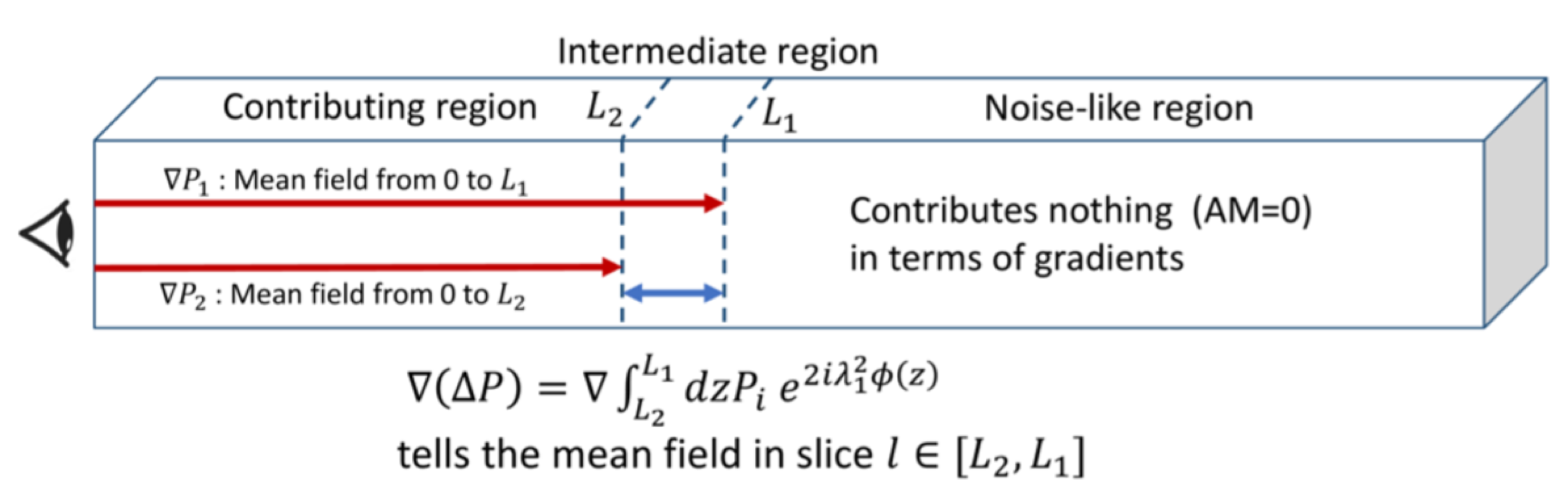}}
\caption{Schematic diagram of the principle of reconstructing the 3D magnetic field structure using SPDGs. The spatial gradient of polarization intensity $\nabla(\triangle P)$ at two neighboring wavelengths $\lambda$ reflects the mean magnetic field structure between the slice $[L_2, L_1]$ (Lazarian \& Yuen, \citeyear{LY2018}).
}
\label{Fig:3Dmap}
\end{figure*}

\section{Synchrotron Radiative Processes in MHD turbulence}\label{SynTheo}

Magnetic field and relativistic electrons, as two key factors, play a decisive role in emitting synchrotron radiation (Ginzburg \& Syrovatskii, \citeyear{Ginzburg_Syrovatskii1965}; Ginzburg, \citeyear{Ginzburg1981}). Assume that relativistic electrons display a power-law spectrum
\begin{equation}
N(E)dE=KE^{-p}dE, \label{eq:distribution}
\end{equation}
where $N$ represents the number density of relativistic electrons with the energy between $E$ and $E+dE$, $K$ the normalization constant, and $p$ spectral index of the electrons. In an environment containing regular and random magnetic fields, synchrotron emissivity can be expressed as the combination of homogeneous ($j_{\rm h}$) and random ($j_{\rm r}$) parts (Crusius \& Schlickeiser, \citeyear{Crusius_Schlickeiser1986})
\begin{equation}
j(\nu, \theta)=\chi j_{\rm h}(\nu, \theta)+\frac{1}{2}(1-\chi)j_{\rm r}(\nu),
\end{equation}
where $\chi$ and $1-\chi$ are the percentage of homogeneous and random parts, respectively. $\theta$ is the angle between field lines and the direction of emission, and the factor $\frac{1}{2}$ denotes half of the power $j_{\rm r}(\nu)$ measured in the each of two polarization directions for turbulent magnetic fields.

\subsection{Radiation in homogeneous Magnetic Field}
In a completely homogeneous field, the synchrotron emissivity of relativistic electrons is expressed by (Ginzburg \& Syrovatskii, \citeyear{Ginzburg_Syrovatskii1965})
\begin{equation}
j_{\rm h}(\nu, \theta)=\int dE N(E) \mathcal{F}(E, \nu, \theta). \label{eq:jh}
\end{equation}
The integrand $\mathcal{F}$, representing the power per frequency unit emitted by a single electron of given energy and pitch angle, is given by (see Rybicki \& Lightman, \citeyear{rybicki1979} for more details on the derivation) 
\begin{equation}
\mathcal{F}(E, \nu, \theta)=\frac{\sqrt{3}{e^3}B\sin{\theta}}{m_{\rm e}c^2}[\frac{\nu}{\nu_{\rm c}}\int_{\nu/\nu_{\rm c}}^{\infty}K_{5/3}(t)dt], \label{P_nu_theta}
\end{equation}
where $K_{5/3}(t)$ is the modified Bessel function of order $5/3$, $f(\frac{\nu}{\nu_{c}})=\frac{\nu}{\nu_{c}}\int_{\nu/\nu_{c}}^{\infty}K_{5/3}(t)dt$ is often referred to as a kernel function, and  $\nu_{c}=3E^2eB{\rm sin}\theta/4\pi m_{\rm e}^3c^5$ is a characteristic frequency of synchrotron emission. 

The power per frequency of a single electron, $\mathcal{F}$, is divided into the perpendicular and parallel components with respect to the magnetic field $\bf{B}$ (Rybicki \& Lightman, \citeyear{rybicki1979})
\begin{equation}
\mathcal{F}_{\perp}(E, \nu, \theta)=\frac{\sqrt{3}e^{3}B\sin{\theta}}{2m_{\rm e}c^{2}}[f(\frac{\nu}{\nu_{c}})+g(\frac{\nu}{\nu_{c}})],\label{eq:Fperp}
\end{equation}
\begin{equation}
\mathcal{F}_{\parallel}(E, \nu, \theta)=\frac{\sqrt{3}e^{3}B\sin{\theta}}{2m_{\rm e}c^{2}}[f(\frac{\nu}{\nu_{c}})-g(\frac{\nu}{\nu_{c}})],\label{eq:Fpara}
\end{equation}
where $g(\nu/\nu_{c})=\frac{\nu}{\nu_{c}}K_{\frac{2}{3}}(\frac{\nu}{\nu_{c}})$ and $K_{\frac{2}{3}}$ is the modified Bessel function of order $2/3$. As a result, the degree of linear polarization can be calculated by
\begin{equation}
\varPi(\nu)=\frac{\mathcal{F}_{\perp}-\mathcal{F}_{\parallel}}{\mathcal{F}_{\perp}+\mathcal{F}_{\parallel}}=\frac{g(\nu/\nu_{c})}{f(\nu/\nu_{c})} \label{degree}
\end{equation}
for a single electron radiation. Replacing $\mathcal{F}$ in Eq. (\ref{eq:jh}) with $\mathcal{F}_{\parallel}$ or $\mathcal{F}_{\perp}$, we can obtain the perpendicular ($j_{\perp}$) and parallel ($j_{\|}$) components of synchrotron emissivity for electron polulations. Hence, we have the degree of linear polarization for the electron populations
\begin{equation}
\varPi(\nu)=\frac{j_{\perp}-j_{\parallel}}{j_{\perp}+j_{\parallel}}. \label{degree1}
\end{equation}
After integration, the total emissivity (Eq.\ref{eq:jh}) can be simplified as
\begin{equation}
j_{\rm h}(\nu, \theta)\propto (B_\perp)^{(p+1)/2}\nu^{-(p-1)/2}, \label{eq:jh1}
\end{equation}
where the spectral index of photons is $\alpha=\frac{p-1}{2}$ and $B_{\perp}=B\sin{\theta}$.

\subsection{Radiation in Random Magnetic Field}
In a completely random field, the calculation of synchrotron emissivity will be more complex compared to a completely homogeneous field, due to the presence of a stochastic direction. The synchrotron emissivity is formulated as (Crusius \& Schlickeiser, \citeyear{Crusius_Schlickeiser1986})
\begin{equation}
j_{\rm r}(\nu)=\frac{1}{4\pi}\int_{0}^{2\pi}d\varphi\int_{0}^{\pi}d\theta \sin{\theta}j_{\rm h}(\nu, \theta), \label{eq:jr}
\end{equation}
by averaging the total spontaneously emitted power for the homogeneous field $j_{\rm h}(\nu, \theta)$ over all possible values of the polar ($\theta$) and azimuthal ($\varphi$) angles. This equation implies that randomness of the magnetic field is maintained ranging from the Larmor radii of the radiating electrons to the size of the emitting source. 

Defining $x=\nu{\rm sin}\theta/\nu_{\rm c}$, Eq. (\ref{eq:jr}) can be rewritten as  
\begin{equation}
j_{\rm r}(\nu)\propto B\int_{0}^{\infty}dEN(E)R(x),
\end{equation}
where $x=\frac{\nu}{(c_{1}BE^{2})}$, $c_{1}=\frac{3e}{4\pi m_{\rm e}^{3} c^{5}}$ and $R(x)=\frac{x}{2}\int_{0}^{\pi}d{\theta}\sin{\theta}\int_{x/\sin{\theta}}^{\infty} K_{5/3}(t)dt$ is associated with the Bessel function of order $5/3$. The analytical results demonstrated that $j_{\rm r}$ remains the same power-law relation for frequency as that of $j_{\rm h}$ (Crusius \& Schlickeiser, \citeyear{Crusius_Schlickeiser1986}). Different from the situation of homogeneous field, $j_{\rm r}$ is related to the total magnetic field strength $B$ and not just its perpendicular component $B_\perp$ in $j_{\rm h}$ expression.

\subsection{Synchrotron self-absorption}

When involving in the synchrotron self-absorption, one can adopt Eq. (\ref{P_nu_theta}) to obtain the following the self-absorption coefficient
\begin{equation}
\kappa(\nu, \theta)=\frac{c^2}{8\pi\nu^2}\int E^{2}\frac{d}{dE}(\frac{N(E)}{E^2})\mathcal{F}(E, \nu, \theta)dE.
\end{equation}
We can thus write total radiation intensity by the radiative transfer equation
\begin{equation}
I(\nu)=\frac{j}{\kappa}(1-e^{-\tau}), \label{I_tau}
\end{equation}
where $\tau=\kappa\ell$ is the absorption optical depth, and $\ell$ is the size of the emitting region.

For the electrons with a power-law distribution, the absorption coefficient can be reduced to
\begin{equation}
\begin{aligned}
\kappa(\nu, \theta)=\frac{p+10/3}{p+2}\mu(\nu, \theta)=G(p)\frac{e^3}{2\pi m_{\rm e}}\\
(\frac{3e}{2\pi m_{\rm e}^3 c^5})^{p/2}K(B\sin\theta)^{(p+2)/2}\nu^{-(p+4)/2},
\end{aligned}
\end{equation}
where $\mu(\nu, \theta)=\frac{p+2}{p+10/3}\kappa(\nu, \theta)$, and
the coefficient $G(p)=\frac{\sqrt{3}}{4}\Gamma{(\frac{3p+2}{12})}\Gamma{(\frac{3p+22}{12})}$ is only associated with the electron exponent. 

To obtain polarization properties, we here introduce the following coefficients in two mutually perpendicular directions 
\begin{equation}
\kappa_{\perp}(\nu, \theta)=\kappa(\nu, \theta)+\mu(\nu, \theta),
\end{equation}
\begin{equation}
\kappa_{\parallel}(\nu, \theta)=\kappa(\nu, \theta)-\mu(\nu, \theta).
\end{equation}
By adapting Eq. (\ref{I_tau}), we have radiation intensity components 
\begin{equation}
I_{\perp}(\nu)=\frac{j_{\perp}}{\kappa_{\perp}}(1-e^{-\kappa_\perp\ell}),
\end{equation}
\begin{equation}
I_{\|}(\nu)=\frac{j_{\|}}{\kappa_{\|}}(1-e^{-\kappa_\|\ell}).
\end{equation}
As for an optically thin medium, i.e., $\kappa_{\perp, \parallel}\ell \ll 1$, the degree of linear polarization is consistent with Eq. (\ref{degree1}).
In the case of optically thick medium, i.e., $\kappa_{\perp, \parallel}\ell \gg 1$,  
\begin{equation}
\Pi(\nu)={\frac{{j_{\perp}/\kappa_{\perp}-j_{\parallel}/\kappa_{\parallel}}}{j_{\perp}/\kappa_{\perp}+j_{\parallel}/\kappa_{\parallel}}}
\end{equation}
can be used to calculate the degree of linear polarization.

\begin{figure*}[]
\centerline{\includegraphics[width=80mm,height=85mm,angle=90]{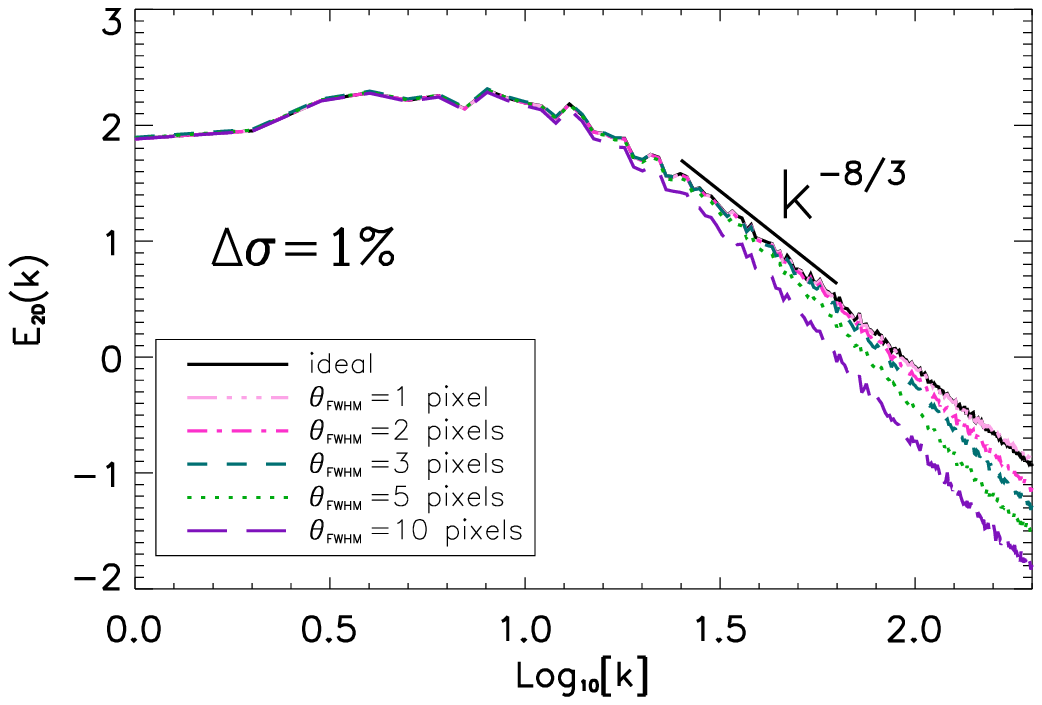} \includegraphics[width=80mm,height=85mm,angle=90]{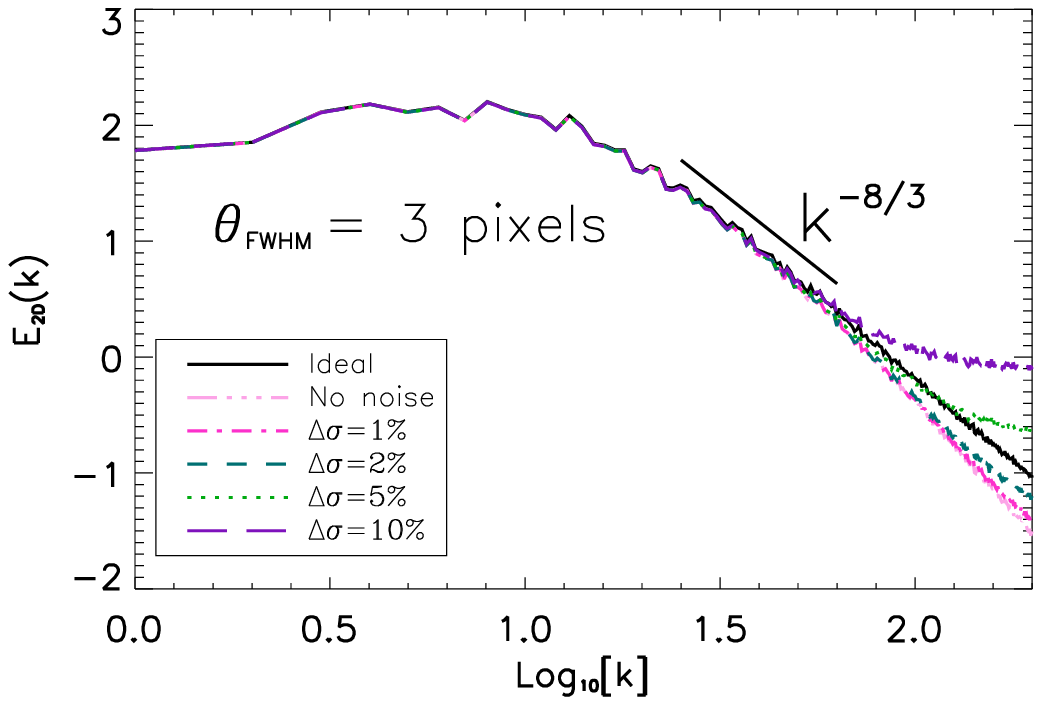}}
\caption{The influences of telescope angular resolution (left panel) and Gaussian noise (right panel) on the power spectrum calculated at the wavelength $\lambda=2.4$ in code units. The symbol $\Delta\sigma$ indicating standard deviation of Gaussian noise accounts for a fraction of the mean synchrotron polarization intensity. The symbol $\theta_{\rm FWHM}$ denoting an effective Gaussian beam is used to convolve 2D maps of Stokes parameters. The original (ideal) slope without considering the resolution and noise is shown in the solid line (Zhang et al. \citeyear{Zhang2018}).
 }  \label{figs:ang_noise}
\end{figure*}

\section{Statistics of Synchrotron Fluctuations} \label{SSRF}

On the basis of synchrotron emissivity in Section \ref{SynTheo}, we can obtain observable synchrotron intensity ($I$) and polarization intensity ($P$) by integrating along the line of sight (Burn, \citeyear{Burn1966}; Waelkens et al., \citeyear{Waelkens2009})
\begin{equation}
I({\bf X})=\int_{0}^{L}(j_{\perp}+j_{\parallel})dz,
\end{equation}
\begin{equation}
{\bf P}({\bf X}, \lambda^{2})=\int_{0}^{L}(j_{\perp}-j_{\parallel})e^{2i\psi({\bf X}, z)}dz,
\end{equation}
where $L$ is the spatial scale of radiative propagation, $\bf X$ is a 2D vector on the plane of the sky and $z$ is the distance variable along the line of sight. The phase (polarization) angle $\psi({\bf X}, z)=\psi_{0}+\lambda^{2}\Phi$ is associated with the intrinsic polarization angle $\psi_{0}$ and the Faraday rotation measure $\Phi$. This is expressed as  
\begin{equation}
\Phi({\bf X},z)\propto\int_{0}^{z}n_{\rm e}({\bf X}, z') B_\|({\bf X}, z') dz',
\end{equation}
where $n_{\rm e}$ is the thermal electron density, and $B_\|$ is the magnetic field parallel component along the line of sight. Defining the intrinsic polarization intensity density as ${\bf P}_{\rm i}={\bf Q}_{\rm i}+i{\bf U}_{\rm i}$ which is independent of wavelengths (this simplification does not affect the statistical results, as verified in Zhang et al., \citeyear{Zhang2018} ), we can write the polarization intensity ${\bf P}={\bf Q}+i{\bf U}$ as
\begin{equation}
{\bf P}({\bf X}, \lambda^{2})=\int_{0}^{L}dz {\bf P}_{\rm i}({\bf X}, z)e^{2i\lambda^{2}\Phi({\bf X}, z)}. \label{Pii}
\end{equation} 

\subsection{Correlation Analysis of Synchrotron Fluctuations}
According to the commonly used method of correlation analysis, the two-point correlation of the polarized intensities can be expressed as (LP16)
\begin{equation}
\begin{aligned}
\langle {\bf P}({\bf X}_{1}, \lambda_{1}^{2}){\bf P}^{*}({\bf X}_{2} , \lambda_{2}^{2})\rangle=\int_{0}^{L} dz_{1}\int_{0}^{L} dz_{2}\\ \langle {\bf P}_{\rm i}({\bf X}_{1}, z_{1}) {\bf P}_{\rm i}^{*}({\bf X}_{2}, z_{2})e^{2i(\lambda_{1}^{2}\Phi({\bf X}_{1}, z_{1})-\lambda_{2}^{2}\Phi({\bf X}_{2}, z_{\rm 2}))}\rangle,
\end{aligned} \label{PPcor}
\end{equation}
where $\bf P^{*}$ denotes the conjugate of the polarization intensity, $\bf X_{1}$ and $\bf X_{2}$ represent two different spatial positions on the plane of the sky, $\lambda_{1}$ and $\lambda_{2}$ correspond to two wavelengths.
Eq. (\ref{PPcor}) can not only characterize the correlation of the spatial separation but also the dispersion of frequencies. Based on this, LP16 proposed two techniques, namely the polarization spatial analysis (PSA) and the polarization frequency analysis (PFA). 

In analogy with the velocity channel analysis (VCA) technique (Lazarian et al., \citeyear{Lazarian_etal2002}), the PSA correlates the polarization signal at different spatial points of the position-position frequency (PPF) space that has similarities with the position-position volume (PPV) space. At a fixed wavelength, the PSA can be expressed as
\begin{equation}
\begin{aligned}
\langle {\bf P}({\bf X_{1}}){\bf P}^{*}({\bf X_{2}})\rangle=\int_{0}^{L}dz_{1}\int_{0}^{L}dz_{2} e^{2i \bar{\phi} \lambda^{2}(z_{1}-z_{2})}\\ 
\times \langle {\bf P}_{\rm i}({\bf X}_{1}, z_{1}){\bf P}_{\rm i}^{*}({\bf X}_{2}, z_{2})\rangle \\
\times e^{-2\lambda^{4}\langle(\triangle\Phi({\bf X}_{1}, z_{1})-\triangle\Phi({\bf X}_{2}, z_{2}))^{2}}\rangle,\label{Pcorr}
\end{aligned}
\end{equation}
where $\bar{\phi}$ is mean Faraday rotation measure density and $\Delta\Phi$ is the fluctuation of Faraday rotation measure. This represents the correlation of polarization intensity as a function of spatial separation ${\bf R}={\bf X}_{2}-{\bf X}_{1}$ at the same wavelength.

The PFA that is analogous to the velocity coordinate spectrum (VCS) correlates the polarization information along a single line of sight at the same spatial position. At a fixed spatial position, the PFA for different wavelengths is expressed by
\begin{equation}
\begin{aligned}
\langle {\bf P}({\lambda_{1}^{2}}){\bf P}^{*}(\lambda_{2}^{2})\rangle =\int_{0}^{L}dz_{1}\int_{0}^{L}dz_{2}e^{2i\bar{\phi}(\lambda_{1}^{2}z_{1}-\lambda_{2}^{2}z_{2})}\\
\langle {\bf P}_{\rm i}(z_{1}){\bf P}_{\rm i}^{*}(z_{2})
e^{2i(\lambda_{1}^{2}\triangle\Phi(z_{1})-\lambda_{2}^{2}\triangle\Phi(z_{2}))}\rangle, 
\end{aligned}
\end{equation}
which is complementary to the PSA, so this enables a more comprehensive acquisition of the properties of MHD turbulence. 

Another important aspect of the development of the synchrotron fluctuation technique is the construction of the statistics of polarization intensity derivative with respect to the squared wavelength $\lambda^{2}$. The correlation statistics of $dP/d\lambda^2$ is formulated as 
\begin{equation}
\begin{aligned}
\langle\frac{{\bf dP}({\bf X}_{1})}{d\lambda^{2}}\frac{{\bf dP}^{*}({\bf X}_{2})}{d\lambda^{2}}\rangle=\int_{0}^{L}dz_{1}\int_{0}^{L}dz_{2}e^{2i\bar{\phi}\lambda^{2}(z_{1}-z_{2})}\\
\times \langle {\bf P}_{\rm i}({\bf X}_{1}, z_{1}) {\bf P}_{\rm i}^{*}({\bf X}_{2}, z_{2})\Delta\Phi({\bf X}_{1}, z_{1})
\Delta\Phi({\bf X}_{2}, z_{2})\\
\times e^{-2{\lambda}^{2}i(\Delta\Phi({\bf X}_{1}, z_{1})-\Delta\Phi({\bf X}_{2}, z_{2}))}\rangle,
\end{aligned}
\end{equation}
which is more sensitive to the fluctuations of Faraday rotation than the polarization intensity correlation given in Eq. (\ref{Pcorr}). The introduction of this relationship allows for more information such as the spectrum of Faraday rotation to be obtained. In addition to considering the correlation function, one can also use the structure function of polarization intensity to reveal the anisotropy of MHD turbulence (see LP16 for detailed formulation). 

\subsection{Analysis of Power Spectrum}
With the correlation analysis of synchrotron fluctuations, the power spectrum of synchrotron polarization is described through the Fourier transform of the two-point correlation tensor as 
\begin{equation}
{E}_{\bf P}({\bf K})=\frac{1}{(2\pi)^2}\int d{\bf R}e^{-i{\bf K\cdot R}}\langle {\bf P}({\bf X}_{1}){\bf P}^{*}({\bf X}_{2})\rangle,
\end{equation}
where ${\bf R}={\bf X}_{1}-{\bf X}_{2}\propto 1/{\bf K}$ is the spatial lag vector on the plane of the sky. Similarly, the power spectrum of the polarization derivative with respect to $\lambda^{2}$ can be written as 
\begin{equation}
{E}_{\bf dP}({\bf K})=\frac{1}{(2\pi)^2}\int d{\bf R}e^{-i{\bf K\cdot R}}\langle \frac{{\bf dP}({\bf X}_{1})}{d\lambda^2}\frac{{\bf dP}^{*}({\bf X}_{2})}{d\lambda^2}\rangle   
\end{equation}
for analyzing the multi-frequency observations. Once the power spectrum distribution of the observed information is obtained through the above two equations, we can obtain the distribution characteristics of turbulent energy in space, and also understand several properties related to turbulence, such as the scaling slope, energy dissipative and injection scales.

\subsection{Analysis of Anisotropy}
Following LP12, we can define the normalized correction function for an arbitrary observable $\digamma$ as 
\begin{equation}
\xi_{\rm \digamma}({\bf R})=\frac
{{\langle \digamma(\bf X)\digamma(\bf{X+R})\rangle}-{\langle \digamma (\bf X)\rangle}^2}
{\langle \digamma(\bf X)^2 \rangle-{\langle \digamma(\bf X)\rangle}^2},\label{eq:NCF}
\end{equation}
where $\bf{X}$ and $\bf R$ denote a spatial position and a separation between any two points on the plane of the sky, respectively. This is related to the normalized structure function of the observable $\digamma$  
\begin{equation}
\tilde{D}_{\digamma}=2(1-\xi_{\digamma}), \label{eq:NSF}
\end{equation} 
by which we have a multipole expansion
\begin{equation}
{\tilde{M}}_{n}({\bf R})=\frac{1}{2}\int_{0}^{2\pi} e^{-in\varphi}{\tilde D}_\digamma({\bf R}, \varphi) d\varphi,
\end{equation}
where $n$ represents the order of multipole, and $\varphi$ is the polar angle. The order of multipole expansion has following meaning: $n=0$ represents the monopole reflecting the isotropy of statistics, $n=1$ denotes the dipole which is zero due to the rotation symmetry of the structure function, $n=2$ is called the quadrupole moment revealing the anisotropy, and $n=4$ is the octupole moment which has a negligible contribution for the anisotropy. Hence, we define the ratio of the quadrupole moment to the monopole one (termed quadrupole ratio, ref. LP16)  
\begin{equation}
\Re_{\rm Q}=\frac{\tilde{M}_{\rm 2}({\bf R})}{\tilde{M}_{\rm 0}({\bf R})}=
\frac{\int_0^{2\pi} e^{-2i\varphi} \tilde{D}_{\digamma}({\bf R},\varphi)~d{\varphi}}
{\int_0^{2\pi} \tilde{D}_{\digamma}({\bf R},\varphi)~d{\varphi}} \label{eq:9}
\end{equation}
to characterize the degree of anisotropy, the absolute value of which is called the quadrupole ratio modulus. 

Alternatively, the ratio of structure functions in the perpendicular direction to each other can also depict the level of anisotropy, and is given by
\begin{equation}
\Re_{\rm S}=\frac{\langle|\digamma_{\parallel}(x+R_{\parallel})-\digamma_{\parallel}(x)|^2\rangle}{\langle|\digamma_{\perp}(y+R_{\perp})-\digamma_{\perp}(y)|^2\rangle},
\end{equation}
where $R=(R_{\parallel}^2+R_{\perp}^2)^{1/2}$ is the spatial separation on the plane of the sky. When $\Re_{\rm S}=1$, statistics represents an isotropy while any deviation from $\Re_{\rm S}=1$ implies an appearance of anisotropy.

\subsection{Statistics of Spatial Gradients}
Combining the local alignment theory of MHD turbulence (GS95; LV99) with the synchrotron polarization fluctuation predictions (LP16), the gradient techniques were proposed to measure the properties of MHD turbulence (Lazarian et al., \citeyear{Lazarian_etal2017}; Lazarian \& Yuen, \citeyear{LY2018}). This is analogous to velocity gradient techniques (Gonz\'alez-Casanova \& Lazarian, \citeyear{GL_2017ApJ}; Yuen \& Lazarian, \citeyear{YL_2017}). The feasibility lies in that magnetic field and velocity fluctuations enter symmetrically, according to the theory of Alfv{\'e}nic turbulence. In the framework of GS95 MHD turbulence theory, it is expected that the alignment of the magnetic field gradient direction is perpendicular to the local magnetic field direction with the eddies (for other different anisotropy models, the sensitivity to this alignment effect may significantly depend on the scales). Since the synchrotron fluctuations are elongated in the direction of the mean magnetic fields, the magnetic fields can be traced by the direction of the spatial gradient of the observed signal.

When involved in Faraday depolarization of synchrotron radiation fluctuations, synchrotron polarization gradients (SPGs) and synchrotron polarization derivative gradients (SPDGs) can reveal more detailed information on the turbulent magnetic field. In particular, the latter can sample the spatial regions close to the observer with the sampling depth controlled by the radiation wavelength (see Figure \ref{Fig:3Dmap}), which provides a new approach to map the 3D distribution of Galactic magnetic fields. 

Specifically, the decorrelation of the Faraday rotation measure is introduced as (LP16; Zhang et al., \citeyear{Zhang2020}) 
\begin{equation}
\label{eq:oneradcon}
\lambda^2\varPhi = 0.81\lambda^2 \int_0^{L_{\rm eff}} dz n_{\rm e} B_\| = 2\pi,
\end{equation}
where $L_{\rm eff}$ is an effective spatial depth of the Faraday rotation sampling. We can thus write the ratio of the depth sampled by the Faraday rotation to the emitting region size as
\begin{equation}
\label{eq:el}
\frac{L_{\rm eff}}{L} \approx \frac{2\pi}{\lambda^2L} \frac{1}{\phi}, 
\end{equation}
with $\phi= {\rm max}(\sqrt{2} \sigma_\phi,\bar{\phi})$. Here, $\sigma_\phi$ denotes the root mean square and $\bar{\phi}$ Faraday rotation density. According to Eq. (\ref{eq:el}), the strong and weak depolarization can be characterized by $L_{\rm eff}/L<1$ and $L_{\rm eff}/L>1$, respectively. In the case of the strong depolarization, the entire space is divided to two different parts shown in Figure \ref{Fig:3Dmap}, from which we can know that only the part of $z<L_{\rm eff}$ experiences the depolarization while the part of $z>L_{\rm eff}$ cannot contribute to the polarization correlation. When observing frequency interval $\triangle\nu=c/\lambda^2\triangle\lambda$ is small enough, the difference of ${\bf P}(\lambda_{1})$ and ${\bf P}(\lambda_{2})$ in the range of $[L_{2}, L_{1}]$ is written as 
\begin{equation}
\triangle {\bf P}\approx\int_{L_{2}}^{L_{1}}dz {\bf P}_{\rm i}({\bf X}, z)e^{2i\lambda^2\Phi({\bf X}, z)},
\end{equation}
from which the expression of the SPDGs is given by
\begin{equation}
\nabla \frac{d|{\bf P}|}{d\lambda^2}\sim \lambda^{-2}\nabla|{\bf P}(\lambda_{1})-{\bf P}(\lambda_{2})|.
\end{equation}

From another perspective, namely, based on the spatial gradient analysis of Stokes parameters $Q$ and $U$ on the complex plane, Gaensler et al. (\citeyear{Gaensler2011}) proposed the synchrotron polarization gradients to constrain sonic Mach number in the warm or ionized ISM. After a simple mathematical derivation, the spatial gradient of polarization intensity can be expressed as 
\begin{equation}
\vert\nabla{P}\vert=\sqrt{(\frac{\partial Q}{\partial x})^2+(\frac{\partial U}{\partial x})^2+(\frac{\partial Q}{\partial y})^2+(\frac{\partial U}{\partial y})^2}.
\label{eq:Pv}
\end{equation}
Subsequently, considering rotational and translational invariance of coordinates, various diagnostic quantities were developed in Herron et al. (\citeyear{Herron2018a}).

\begin{figure} 
\begin{center}
\includegraphics[width=70mm,height=85mm,angle=90]{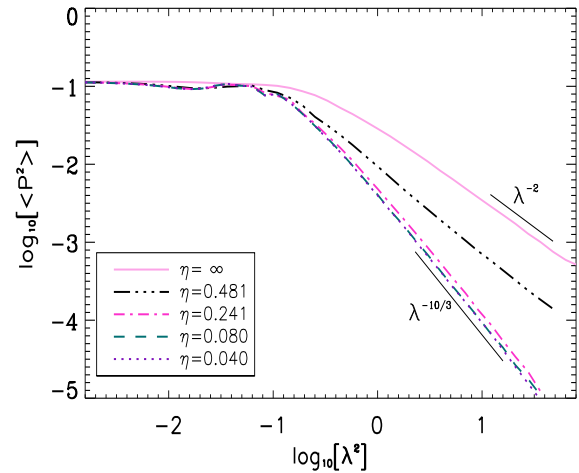}
\caption[ ]{The transition of polarization variance from the stochastic Faraday rotation fluctuation dominant regime for $\eta=\infty$, to the uniform fluctuation dominant one for $\eta=0.080$ in 3D synthetic simulation with $4096^3$ resolution. 
The underlying MHD turbulence corresponds to Komolgorov spectrum of the index $11/3$ (Zhang et al., \citeyear{Zhang2016}). } \label{fig:var3dmean} \end{center}
\end{figure}

\section{Application of Statistical Techniques}
\subsection{Measurement of Scaling Slope}
LP12 predicted that the power spectrum, correlation, and structure functions of synchrotron radiation fluctuations, termed two-point statistics, can recover the scaling slope of turbulent magnetic fields. Numerical simulations show that the correlation function does not work in revealing scaling slopes, but it can be used to explore the correlation scale and the effect of the electron spectral index on turbulence (Herron et al., \citeyear{Herron2016}; Lee et al., \citeyear{Lee2016}; Zhang et al., \citeyear{Zhang2016}). The main reason we think is that LP16 performed theoretical studies under the condition of the power-law correlation model, from which they predicted the possibility of obtaining the scaling slope for the corresponding model. The correlation function itself, however, has a divergent behavior for numerical simulation or observational data, because Stokes parameters $Q$ and $U$ contain the information of negative values arising from the change in the direction of the turbulent magnetic field.

Numerical results demonstrated that the scaling slope of the underlying MHD turbulence can be recovered approximately by the two-point structure function of synchrotron radiation fluctuations (Lee et al., \citeyear{Lee2016}; Zhang et al., 2022). From the perspective of numerical simulations, the direct calculation of the spatial structure function is more time-consuming, compared with the correlation function or power spectrum. Alternatively, one can first get the statistics of the correlation function, provided that it maintains convergence, and then calculate the structure function using the relation of ${\rm SF}(\bf{R})=2[{\rm CF}(0) - {\rm CF}(\bf{R})]$. In general, the two-point structure functions are unable to correctly reproduce the scaling of a spectrum with a slope $\sim -3$  or steeper (sometimes problems may arise already at slopes slightly shallower than $-3$, if the underlying spectrum is not a perfect power law). For instance, Lee et al. (\citeyear{Lee2016}) measured the scaling slope of 5/3 with the change in a numerical resolution. It was pointed out that the scaling slope of the two-point structure function, calculated directly from 2D synthetic data, is still less than 5/3 expected for the Kolmogorov spectrum, even with $8192^2$ resolution. Thus, using a two-point structure function to extract the scaling slope of turbulence requires more precise statistics with higher numerical resolution. Alternatively, the multi-point structure functions are recommended to achieve better convergence of the results (see Cho \& Lazarian, \citeyear{Cho_Lazarian2009}; Cho, \citeyear{Cho2019}).

Obtaining the scaling slope through the power spectrum of synchrotron fluctuations should be one of the most efficient ways, as the calculation of the power spectrum invokes the fast Fourier transform in the wave-number space to effectively accelerate the process. Adopting both synthetic and MHD turbulence simulation data, Lee et al. (\citeyear{Lee2016}) obtained the power spectrum and its derivative with respect to the wavelength of synchrotron polarization arising from synchrotron polarization radiation together with Faraday rotation fluctuations. This study succeeded in obtaining the scaling slope and demonstrated that simulations are in agreement with the theoretical prediction in LP16. In addition, using simulated interferometric observations on how to recover the scaling slope was also attempted. Notice that the study mentioned above is based on the spatially coincident synchrotron emission and Faraday rotation regions. Furthermore, considering spatially separated and compounded synchrotron radiation and Faraday rotation fluctuations, Zhang et al. (\citeyear{Zhang2018}) studied how to extract the scaling slope of synchrotron polarization intensities. In the short wavelength range, the power spectra reveal fluctuation statistics of the perpendicular component of turbulent magnetic fields, and the spectra reflect the fluctuation of the Faraday rotation in the long-wavelength range. As shown in Figure \ref{figs:ang_noise}, the effects of telescope angular resolution and noise structure do not hinder the use of power spectrum methods to obtain scaling slopes of MHD turbulence in the large-scale inertial range.   

\begin{figure*}[t]
\centerline{\includegraphics[width=0.90\textwidth,height=0.30\textheight]{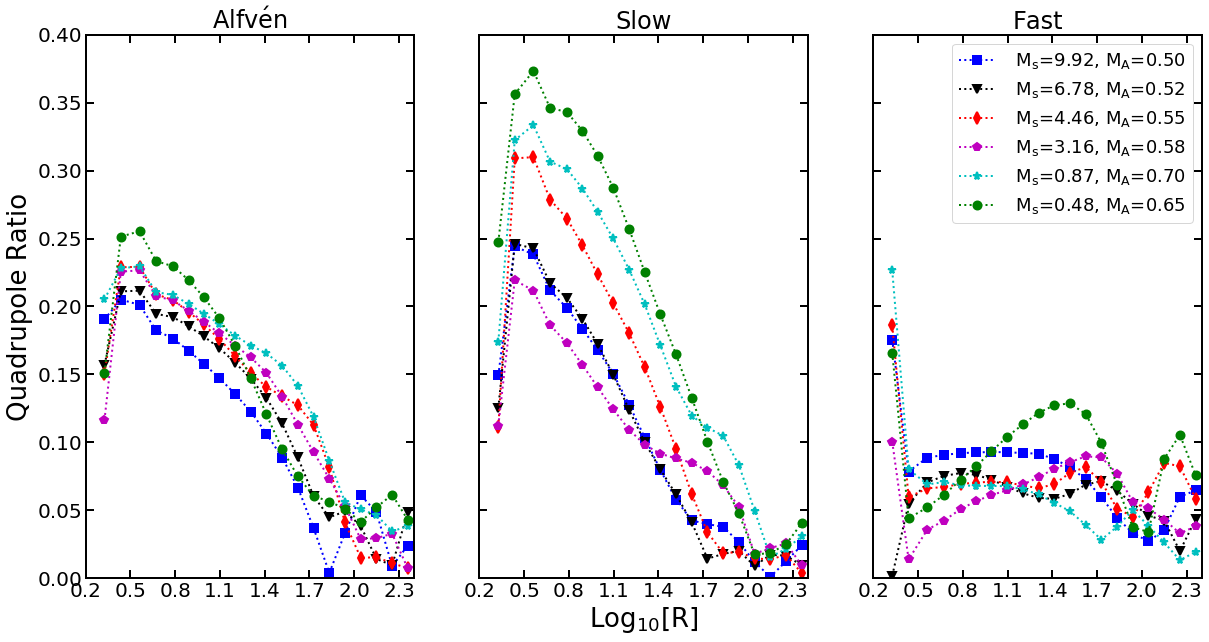}}
\caption{Quadrupole ratio of synchrotron polarization intensities for Alfv\'en, slow and fast modes as a function of radial separation of maps in sub-Alfv\'enic regime (Wang et al., \citeyear{Wang2020}).
}
\label{Fig:QRMaMs}
\end{figure*}

Motivated by one-point statistics in LP16 (termed PFA), which is characterized by the variance of polarized emission as a function of the squared wavelength along a single line of sight, Zhang et al. (\citeyear{Zhang2016}) studied how to extract the scaling slope of underlying MHD turbulence. To depict the level of turbulence, they defined a ratio $\eta$ of the standard deviation of the line-of-sight turbulent magnetic field to the line-of-sight mean-magnetic field. As shown in Figure \ref{fig:var3dmean}, a large ratio ($\eta\gg1$) characterizing a turbulent field dominated regime reflects the polarization variance $\left<P^2\right>\propto\lambda^{-2}$, and a small ratio ($\eta\lesssim0.2$) characterizing a mean-field dominated regime reveals the polarization variance $\left<P^2\right>\propto\lambda^{-2-2m}=\lambda^{-10/3}$ with $m=2/3$ for the Kolmogorov scaling. As a result, the turbulent spectral index was successfully recovered by the polarization variance in the case of the mean-field dominated regime, i.e., a small $\eta$. At the same time, it was pointed out that the change in cosmic ray electron spectral indices cannot affect the scaling slope measurement of magnetic turbulence. This provides a new method for recovering the scaling index of MHD turbulence and agrees with the theoretical prediction of LP16. Interestingly, this technique has been successfully applied to explain the depolarization in the optical wavebands from the active galactic nucleus (Guo et al., \citeyear{Guo_2017}).

\subsection{Measurement of Anisotropy}
As described in Section \ref{thoer}, the prediction of the anisotropy (eddy's elongation along the local magnetic field direction) in turbulent cascade is an important contribution of GS95. It is challenging to test the {\it local} anisotropy of MHD turbulence from an observational point of view because the observational signal is inevitably involved in the integration along the line of sight in the observer's frame of reference. Note that integrated synchrotron fluctuations along the line of sight were predicted by LP12 to be anisotropic, with the eddy's major axis aligned with the direction of the mean magnetic field instead of the local field within the MHD turbulence volume. 

The spatial second-order structure function of radiation fluctuations is a very direct way to reveal the existence of anisotropy in MHD turbulence, although the anisotropy here does not directly correspond to the local anisotropy favored by the modern MHD turbulence theory. If visualizing a contour map for 2D structure function of observations, one can easily see the anisotropic distribution, from which the major axis direction of the elongated structure will qualitatively characterize the mean magnetic field orientation (see Figure 1 in Wang et al. (\citeyear{Wang2020}) for an example). Considering the ratio of two measurement directions perpendicular to each other from observed polarization intensities, called an anisotropic coefficient, we can quantitatively determine the degree of the anisotropy of the MHD turbulence in the global frame of the reference; refer to Lee et al. (\citeyear{Lee2019}) and Wang et al. (\citeyear{Wang2020}) for more details.

A new more-precise method to measure the anisotropy is termed as the quadrupole ratio described in Section \ref{SSRF}. This method was proposed in LP12 that derives many analytical formulae including the predictions from the Alfv\'en, slow and fast modes in the high or low $\beta$ plasma regime. A numerical test of theoretical predictions on the quadrupole ratio was attempted by Herron et al. (\citeyear{Herron2016}) based on MHD turbulence simulation. We have to point out that their testing used a simplified version rather than a general formula. In particular, they adopted the compressible MHD turbulence data to study the isotropic version of formulae, which fails to provide proof of the usability of the analytical formula. Considering the angle change in between the mean magnetic field direction and the line of sight, Wang et al. (2022) recently tested the theoretical prediction of LP12 in detail and confirmed the correctness of the analytical formulae, providing convenience to study the anisotropy of MHD turbulence by using the analytical expressions.

How the anisotropic structure changes with observation frequency were studied in Lee et al. (\citeyear{Lee2019}) by the quadrupole ratio. Their studies showed that in the high-frequency range, the observed polarization exhibits the averaged structures of both foreground and background regions, while in the low-frequency range, the large-scale structures are wiped out by a strong Faraday depolarization and the small-scale anisotropy reflects magnetic field structure from the foreground region. 

By decomposing the compressible MHD turbulence into three plasma modes, i.e., Alfv\'en, slow and fast modes, Wang et al. (\citeyear{Wang2020}) demonstrated that the quadrupole ratio statistics of the polarization intensity from the Alfvén and slow modes present a significant anisotropy, while the fast mode statistics show isotropic structures (see Figure \ref{Fig:QRMaMs}). Those findings are in agreement with the earlier direct numerical simulations provided in CL02. As a result, the quadrupole ratio method can be applied to the measurement of the anisotropy of MHD turbulence in various astrophysical environments. 

\subsection{Measurement of Magnetic Field Orientations}
The traditional method of measuring the magnetic field direction is called the synchrotron polarization vector, which is based on the fact that the observed polarization vectors are perpendicular to the magnetic field directions within the emitting source. Note that the limitation of this method cannot be applied to observations with Faraday rotation.

Using synchrotron intensity gradients (SIGs) together with both MHD turbulence simulation and PLANCK archive data, Lazarian et al. (\citeyear{Lazarian_etal2017}) demonstrated the practical applicability of the gradient technique described in Section \ref{SSRF}, comparing the measured directions of magnetic fields arising from gradient technique with that of the traditional polarization vector method. Furthermore, SIGs were advanced in Lazarian \& Yuen (\citeyear{LY2018}) to SPGs and SPDGs. Subsequently, to explore the magnetic field measurement capabilities of synchrotron gradient techniques, SPGs and SPDGs were generalized and applied to a variety of possible astrophysical environments in a series of studies including different turbulence regimes, the gradient of various diagnostic quantities, realistic observations of Galactic diffuse media, as well as spatially separated synchrotron radiation and Faraday depolarization regions.

\subsubsection{Gradient Measurement in Various Turbulence Regimes} 
Based on synthetic observations from MHD turbulence simulations, Zhang et al. (\citeyear{Zhang2019a}) extended both SPGs and SPDGs to super-Alfv\'enic regimes and studied how to trace the directions of turbulent magnetic fields. Focusing on multifrequency measurements in the presence of strong Faraday rotation, they provided the procedures of how to recover the projected mean magnetic fields on the plane of the sky and the local magnetic fields within a tomographic slice. The results demonstrate that in the low-frequency strong Faraday rotation regime, the SPGs have a significant advantage over the traditional polarization vector method in tracing projected mean magnetic fields and the SPDGs are applicable to tracing the local magnetic field directions in the local frame of reference of MHD turbulence. At the same time, their parameter research provides a testing ground for the application of the new techniques to a large number of data cubes, such as those from LOFAR and SKA.

\subsubsection{Gradient Measurement of Various Diagnostic Quantities}

The various diagnostic quantities, including the polarization directional curvature, polarization wavelength derivative, and polarization wavelength curvature, are derived in Herron et al. (\citeyear{Herron2018a}) considering the translational and rotational invariance of the observed quantities. It is noticed that the authors were not aware of the relation between the magnetic field direction and gradients of synchrotron radiation gradients, and did not explore whether the new diagnostics they introduced can trace magnetic field in observations.

The spatial gradients of various diagnostics provided in Herron et al., (\citeyear{Herron2018a}) have been proposed to trace the directions of the mean magnetic field in Zhang et al. (\citeyear{Zhang2019b}), who found that the gradients of various diagnostics can effectively determine the direction of the projected magnetic fields on the plane of the sky. One of these techniques, namely the maximum of the radial component of the polarization directional derivative, is the best robust method for tracing the mean magnetic field directions in the Galactic ISM.

The synergy of various diagnostic gradients is complementary in tracing the actual direction of interstellar magnetic fields, especially in the low-frequency Faraday rotation regime where the traditional polarization vector method cannot work. Applying the synergies of synchrotron diagnostic gradients to the archive data from the Canadian Galactic Plane Survey (Taylor et al., \citeyear{Taylor2003}), Zhang et al. (\citeyear{Zhang2019b}) showed that various diagnostic techniques make consistent predictions for the Galactic mean magnetic field directions.

\subsubsection{Gradient Measurement of Spatially Separated Regions} 
When confronted with realistic observations, the synchrotron polarization radiation obtained by the observers could originate from complex turbulence regions rather than a single spatially coincident radiation and Faraday rotation region. Can the synchrotron polarization gradient still measure the direction of the magnetic field for more complex astrophysical settings?

The measurement capabilities of the SPGs were studied in Wang et al. (\citeyear{Wang2021}) through constructing spatially separated models, focusing on exploring how the Faraday rotation density in the foreground region affects the measurement of the projected magnetic field. Their numerical results showed that the use of the SPG technique for a complex astrophysical environment can successfully trace the projected mean magnetic field direction within the emitting-source region independent of radiation frequency. 

\subsubsection{Measurement of the Local Magnetic Field Direction} 
Thanks to the inevitable accumulation of the observed signal along the line of sight, i.e., less locality, it is extremely challenging to measure the local magnetic field in MHD turbulence. However, the measurement of the local magnetic field will be direct evidence to verify the modern MHD turbulence anisotropy theory since this theory is based on the local magnetic field reference system (see Section \ref{scaling_power}). At the same time, the acquisition of the local magnetic field information is a prerequisite for reconstructing the Galactic 3D magnetic field as well. 

Using gradient statistics of synchrotron polarization derivative with respect to the squared wavelength $dP/d\lambda^2$, the measurement of the local magnetic field direction has been implemented in Zhang et al. (\citeyear{Zhang2020}) based on data cubes obtained with MHD turbulence simulations. They demonstrated that in the low-frequency strong Faraday rotation regime, the statistic analysis of the spatial gradient of $dP/d\lambda^2$ can indeed reveal the local magnetic field direction and the local anisotropy of the underlying MHD turbulence which increases with increasing the radiation frequency.

As illustrated in Figure \ref{Fig:3Dmap}, this technique can successfully extract the local information of MHD turbulence in the tomography space along the line of sight by using multi-frequency observations at adjacent frequencies. Unlike the traditional Faraday rotation synthesis (e.g., Burn, \citeyear{Burn1966}), the polarization gradient techniques are not affected by the Faraday rotation and can improve the 3D model of the Galactic magnetic field effectively.

\subsection{Measurement of Magnetization Strength}
Several statistical methods, such as Genus, Tsallis, skewness, and kurtosis statistics, are sensitive to the magnetization level $M_{\rm A}$ of MHD turbulence (e.g., Burkhart et al., \citeyear{Burkhart2012}; Herron et al., \citeyear{Herron2016}). In addition, the correlation function (e.g., Lazarian et al., \citeyear{Lazarian_etal2002}; Burkhart et al., \citeyear{Burkhart2014}; Esquivel et al., \citeyear{Esquivel2015}) and structure function (Hu et al., \citeyear{Hu2021}; Xu \& Hu, \citeyear{Xu2021}) analysis are also suggested as alternative techniques of measuring $M_{\rm A}$. At present, these studies are based on velocity information analysis from MHD turbulence simulations. Similarly, the adoption of synchrotron fluctuations should be on the agenda. 

In analogy with velocity gradient techniques, Lazarian et al. (\citeyear{Lazarian2018}) first discussed that synchrotron gradient techniques should be able to measure the magnetization level. The feasibility of the two methods was explored in Carmo et al. (\citeyear{Lorena2021}) to obtain the level of magnetization from synchrotron polarization fluctuation, namely the top-base and the circular standard deviation. They claimed that the signal-to-noise ratio is more severe for the top-base method, but still reliable with the standard deviation method. In short, the finding on the power-law relation between $M_{\rm A}$ and synchrotron gradient statistics is important for determining $M_{\rm A}$ from the current or future available data cubes.     

\section{Discussion}

The progress in understanding the role of magnetic fields for processes in a specific astrophysical environment depends on our ability to obtain their properties from MHD turbulence. Unfortunately, it is notoriously difficult for studying the properties of magnetic fields. In practice, almost every technique presents its limits. For example, the difficulty with dust polarization measurement lies in the inability to measure the magnetic field strength along the line of sight and insufficient dust grain alignment efficiencies at low Galactic latitudes, preventing observers from deriving the information of magnetic fields in the ISM. For another example, the difficulty with Zeeman splitting measurement is that it can only measure the line of sight component strength of the magnetic field, access the high magnitude end of the interstellar magnetic fields in a mildly turbulent environment, and provide limited morphology information (Crutcher, \citeyear{Crutcher2009}). Last but not least, a serious limitation of the well-known Faraday tomography is that the line-of-sight component of the magnetic field in turbulence is expected to change its sign, resulting in the Faraday depth 
to become ambiguous (see Ho et al. \citeyear{Ho2019} for a comparison of measure performance between Faraday tomography and SPGs; refer also to Akahori et al. \citeyear{Akahori2018} for the recent review on other more methods and applications in specific astrophysics). 

The further development of related new techniques can potentially extract other properties of MHD turbulence. In the framework of synchrotron fluctuations, many topics deserve to be further studied. As an example, visualization of the structure or correlation functions of synchrotron diagnostics can present an anisotropic feature, by which one could trace the mean magnetic field orientations and measure magnetization strength. It should be encouraged to explore the feasibility of this technique for measuring magnetic fields, the fundamental spirit of which is similar to the correlation function anisotropy technique on velocity information (Lazarian et al., \citeyear{Lazarian_etal2002}; Esquivel \& Lazarian, \citeyear{Esquivel_Lazarian2005}; Hu et al., \citeyear{Hu2021}; Xu \& Hu, \citeyear{Xu2021}). We would like to emphasize that the synchrotron dispersion techniques discussed currently are also complementary to other techniques of magnetic field studies, such as velocity channel analysis, velocity coordinate analysis, and velocity gradient techniques. The synergistic use of multiple techniques is necessary because one will be able to improve the reliability of magnetic field measurements, by fully utilizing the advantages of each technique.

None of the relevant papers covered in this review consider synchrotron self-absorption effects (see Section \ref{SynTheo} for self-absorption processes). However, when the brightness temperature of synchrotron radiation approaches the `thermal' temperature of emitting relativistic electrons, the synchrotron self-absorption effect is expected to be important in the case of low frequencies. In the framework of new techniques developed, we expect that the effects of self-absorption can provide yet another method of testing the 3D structure of the magnetic field. Indeed, only the regions closer to the observer and less affected by self-absorption are expected to be probed under strong self-absorption. This effect is similar to the effect of the dust self-absorption in spectral line statistics that was explored analytically in Kandel et al. (\citeyear{Kandel2017b}) and numerically in Yang et al. (\citeyear{Yang2021}).

In the recent studies related to synchrotron radiation techniques, one simply assumes a homogeneous, isotropic non-thermal relativistic electron distribution with a power-law energy spectrum. In fact, the specific properties of the relativistic electrons would depend on the cascade process of MHD turbulence, that is, it is also necessary to solve the so-called turbulent heating problem between MHD scales and electron plasma scales. Since part of the initial cascading turbulent energy will be removed by the ion heating/energizing at ion scales before it reaches the electron scales, only a fraction of the energy cascading from the large MHD scales will end up in heating/energizing the electrons which will determine a specific spectral energy distribution $N(E)$ to produce the observed synchrotron radiation. What is this fraction is currently a big open question and answering it would require understanding what happens at turbulent fluctuations during the transition from MHD scales to kinetic scales, as well as the role of various collisionless processes leading to particle heating that could be at work within different plasma regimes (e.g., Quataert, \citeyear{Quataert1998};  Schekochihin et al., \citeyear{Schekochihin2009}; Chandran et al., \citeyear{Chandran2010}; Cranmer, \citeyear{Cranmer2014}; Arzamasskiy et al., \citeyear{Arzamasskiy2019}; Cerri et al., \citeyear{Cerri2019}; Zhdankin et al., \citeyear{Zhdankin2019}; Vasquez et al., \citeyear{Vasquez2020}; Cerri et al., \citeyear{Cerri2021}). As a result, different anisotropy models (i.e., different turbulence theories at MHD scales; see Section \ref{Controv}) would determine different properties of fluctuations at kinetic scales which is essential to determine the differential heating of ions and electrons.

Compared to the Faraday rotation synthesis to obtain the 3D magnetic fields in the galaxies (e.g., Burn, \citeyear{Burn1966}; Beck, \citeyear{Beck2015}), where the Faraday depth is changing its sign along the line of sight, making the determination of the magnetic field component $B_\parallel$ ambiguous (see Ferriere, \citeyear{ferriere2016}), the advantage of the polarization gradient technique is that it is not affected by Faraday rotation and will effectively improve the measurement of our current Galactic 3D magnetic field model. The correct understanding of this model, as a fundamental ingredient, is indeed very important, e.g., for modeling the anisotropic cosmic-ray propagation in the Galaxy (e.g., Cerri et al., \citeyear{Cerri2017JCAP}; Reichherzer et al., \citeyear{Reichherzer2022}) which can explain some anomalies inferred from gamma-ray emission (Acero et al., \citeyear{Acero2016}; Yang et al., \citeyear{Yang2016PhRvD}), and is critical to obtain realistic Galactic evolution (e.g., Beck, \citeyear{Beck2015}).

Furthermore, an accurate understanding of Galactic magnetic fields is indispensable for searching the B modes of the Cosmic Microwave Background (CMB, Zucca et al., \citeyear{Zucca2017}), which is essential for understanding the evolution of the early universe. The CMB B mode studies require more accurate removal of the Galactic foreground contamination, which is achievable provided that the detailed information of Galactic magnetic fields can be obtained (e.g., Cho \& Lazarian, \citeyear{Cho_Lazarian2010}). Therefore, it is promising that using the new synchrotron gradients performs foreground removal for studying the CMB B modes on the plane of the sky. We expect that there are different properties between the Galactic foreground SPGs and cosmological perturbation one, that is, the direction traced by the former is perpendicular to the foreground polarization, while the latter does not follow this relation. In addition, the CMB E and B mode ratio has been demonstrated to be sensitive to the fraction of the fast modes in MHD turbulence and the magnetization strength (Kandel et al., \citeyear{Kandel_CMB2017}). Furthermore, one can obtain the local properties of the CMB E and B modes using the SPDGs, which provides an alternative way of studying the distribution of compressible MHD turbulence modes over the sky and of determining the magnetization level of the Galactic ISM. 

Most of the work that has been carried out is based on numerical simulations, which on the one hand helps to test the feasibility of the technique, and on the other hand, provides the test grounds for studying magnetic fields using a large number of data cubes from the LOFAR and SKA. Hence, the next step is necessary to advance the application of the technique using observational data to test the performance of the technique in a specific astrophysical setting

\section*{Author Contributions}
All authors listed have made a substantial contribution to the work and approved it for publication.

\section*{Funding}
This work was partially supported by the National Natural Science Foundation of China (Grant Nos. 11973035 and 11703020) and by the Hunan Province Innovation Platform and Talent Plan- HuXiang Youth Talent Project (No. 2020RC3045). 

\section*{Acknowledgments}
We would like to thank two anonymous referees for valuable comments and Alex Lazarian for many constructive discussions.

\end{document}